\documentclass[aps,prl,reprint,superscriptaddress]{revtex4-2}

\usepackage[utf8]{inputenc}
\usepackage[T1]{fontenc}
\usepackage{lmodern}
\usepackage{graphicx}
\usepackage{enumerate}
\usepackage[explicit]{titlesec}
\usepackage{blindtext,color}
\usepackage{hyperref}
\usepackage{amsmath}
\usepackage{amssymb}
\usepackage{amsthm}
\usepackage{bm}
\usepackage {tensor}
\usepackage{etoolbox}
\usepackage{enumitem}
\usepackage[english]{babel}
\usepackage{dsfont}
\usepackage{float}
\usepackage{slashed,braket}
\usepackage{stackengine}
\usepackage{mathrsfs}
\usepackage{verbatim}
\usepackage{array}
\usepackage{upgreek}
\usepackage{bbm}
\numberwithin{equation}{section}

\newcommand{\beq}{\begin{equation}}
\newcommand{\eeq}{\end{equation}}
\newcommand{\beqa}{\begin{eqnarray}}
\newcommand{\eeqa}{\end{eqnarray}}

\begin{document}

\title{Intrinsic (valley) thermal Hall conductivity of phonons in graphene for integer quantum Hall phases}

\author{M. Selch}
\affiliation{Bar-Ilan University, Tel Aviv, Israel}


\begin{abstract}
We calculate the intrinsic contribution to the (valley) thermal Hall conductivity of phonons in (Semenoff-gapped) graphene in the quantum Hall regime where bulk electron thermal transport is suppressed. The (valley) thermal Hall transport of phonons considered here originates from a phonon (valley) Hall viscosity induced by electron (valley) Hall conductivity and viscosity via geometric electron-phonon coupling. The so generated phonon (valley) Hall viscosity is identified with the emergent (valley) Hall viscosity recently introduced in \textit{Phys. Lett. A} \textbf{595}, 132096 (2026). While our calculations refer to graphene, they may very well be generalized to other Dirac materials like group-VI transition metal dichalcogenides. We discuss the prospect of measuring the phonon thermal Hall conductivity in graphene-based heterostructures and estimate the corresponding valley response induced by electrons in the inversion symmetry broken phase. 
\end{abstract}

\maketitle
\textit{Introduction.—}
The thermal Hall effect is a powerful probe of geometric and topological properties of quantum matter. Unlike electrical Hall transport, heat may as well be carried by electrically neutral quasiparticles, making the thermal Hall conductivity directly sensitive to a larger variety of possible heat carriers, microscopic interactions and emergent collective phenomena. The potential to observe thermal Hall transport from carriers other than bulk electrons has recently attracted considerable attention. Experimental measurements of sizeable thermal Hall signals predominantly in magnetic insulators where bulk electron thermal transport is strongly suppressed (see \cite{guo2022thermal,shragai2026phonon} and references therein) have stimulated an intensive search for the microscopic origin of transverse heat transport and the identification of the relevant heat carriers.
Our interest here is dedicated to phonon thermal Hall conductivity \cite{guo2022thermal,chen2022large,li2023the,jin2025discovery,dhakal2025theory,lefrancois2022evidence} for whose generation several distinct mechanisms have now been identified. Intrinsic mechanisms rely on the Berry curvature acquired by phonon bands upon breaking time-reversal symmetry \cite{qin2012berry,saito2019berry,ye2021phonon}, giving rise to an anomalous phonon velocity analogous to the electronic anomalous Hall effect \cite{liu2017pseudospins}, while subsequent work demonstrated that charged defect- \cite{flebus2022charged}, impurity-induced \cite{guo2021extrinsic}, resonantly tuned \cite{sun2022large,guo2022resonant} and collective fluctuation \cite{mangeolle2022phonon} skew scattering provide additional microscopic routes to extrinsic transverse phonon transport. The phonon Hall viscosity-induced intrinsic phonon thermal Hall transport \cite{ye2021phonon} directly links the quantum geometry of an electron fluid in an external magnetic field to the dynamics of the crystal lattice through electron-phonon coupling.\par
Hall viscosity is a dissipationless response coefficient characterizing fluids with broken time-reversal symmetry and has been extensively studied in quantum Hall systems as a geometric topological transport coefficient \cite{avron1995viscosity,avron1998odd,read2009non,read2011hall,bradlyn2012kubo,abanov2014electromagnetic,gromov2014density,hoyos2012hall,cho2014geometry,hoyos2014hall} and measured in graphene-based heterostructures in the semiclassical, hydrodynamic regime for electrons via nonlocal transport measurements \cite{berdyugin2019measuring,kim2025viscous}. A phonon Hall viscosity \cite{zhang2010topological,barkeshli2012dissipationless,qin2012berry,ye2021phonon} in graphene may be viewed as arising from emergent gauge and geometric properties inherited from the underlying electron fluid, as we will outline, which generally competes with the intrinsic properties of the lattice itself \cite{flebus2023phonon}. 
The relevance of Hall viscosity-induced phonon thermal transport has recently received remarkable experimental support making this mechanism particularly appealing. Using ultrasonic measurements of the acoustic Faraday effect, Shragai et al. directly measured a finite phonon Hall viscosity in the candidate Kitaev material $\alpha$-RuCl$_3$ and demonstrated that the resulting intrinsic phonon thermal Hall conductivity quantitatively accounts for a substantial fraction of the experimentally observed thermal Hall signal \cite{shragai2026phonon}. These experiments establish phonon Hall viscosity as an experimentally accessible property of lattice vibrations and identify the acoustic Faraday effect as a direct probe of phonon Berry curvature generated through phonon Hall viscosity. Earlier experimental findings in $^3$He-B have been reinterpreted in this light \cite{tuegel2017hall}.\par
Compared with magnetic insulators, considerably less attention has been devoted to (Hall viscosity-induced) phonon thermal Hall transport in Dirac materials. In graphene \cite{castroneto2009the}, thermal Hall transport measurements have thus far been confined to the ultra-low temperature quantum Hall regime, where heat transport of electrons is suppressed and restricted to chiral electronic edge modes and the associated thermal Hall conductivity reflects topologically protected edge-state transport \cite{srivastav2019universal,srivastav2022determination}. 
This circumstance naturally motivates the search for sizable Hall transport mechanisms for phonons operating in the same regime, where the geometric response of the electronic fluid may be transferred to the lattice through electron-phonon coupling and become experimentally observable. Hybridization of inter-Landau level excitations (magnetoexcitons) and $E_{2g}$ optical phonons has already been shown in graphene via magneto-Raman experiments \cite{kim2013measurement} setting the stage for further investigations of electron induced phonon phenomena in the presence of an external magnetic field.\par
Inspired by the developments in magnetic insulators and our recent results \cite{selch2026emergent} on Hall viscosity in graphene(-like systems), which themselves were inspired by \cite{cortijo2015hall,heidari2019hall} suggesting a sizable Hall transport mechanism, in this Letter we investigate the thermal Hall conductivity of acoustic phonons in graphene in the integer quantum Hall regime. We assume that the transverse phonon response originates from the intrinsic phonon Hall viscosity-induced mechanism, though the corresponding extrinsic mechanism \cite{guo2021extrinsic} may be sizable as well. Within this framework we calculate both the conventional thermal Hall conductivity and its valley analogue for substrate-supported Semenoff-gapped graphene.\par
The valley degree of freedom associated with inequivalent Dirac points in the Brillouin zone enables a variety of valley-dependent transport phenomena. This observation has established valleytronics as a major research direction in condensed matter physics, with the valley Hall effect \cite{xiao2007valley,song2014topological,song2016giant} providing one of its most prominent manifestations. In particular, graphene on inversion-symmetry-breaking substrates such as hexagonal boron nitride (hBN), biased Bernal bilayer graphene and group VI transition metal dichalcogenides (TMD's) have emerged as prototypical experimental platforms in which valley currents can be generated, manipulated, and detected through nonlocal transport measurements \cite{gorbachev2014detecting,shimazaki2015generation,yin2022tunable,sui2015gate,mak2014the}.
The geometric origin of the valley Hall effect naturally suggests the existence of corresponding valley-dependent geometric transport coefficients like the recently discussed valley Hall viscosity \cite{selch2026valley} and then a valley thermal Hall conductivity. 
To our knowledge, this work constitutes the first microscopic prediction of a valley phonon thermal Hall effect induced by Hall viscosity (see \cite{zhai2020topological,chen2022magnon} for a discussion of magnon valley thermal Hall transport). Although our explicit calculations are carried out for graphene-based systems, the underlying mechanism relies only on the existence of massive Dirac quasiparticles as well as gauge and vielbein field acoustic phonon coupling and is therefore expected to apply more broadly to other (inversion symmetry-broken) Dirac materials, including monolayer TMD's.\par
Finally, we estimate the magnitude of the predicted thermal Hall signals and compare the responses in the charge and valley channels under the assumption that nearly complete valley polarization can be achieved experimentally in the latter case to isolate the valley response. These estimates identify the parameter regime in which the proposed effect could become observable and provide quantitative guidance for future thermal Hall experiments on graphene-based heterostructures. More generally, our results establish a connection between Hall viscosity, phonon transport, and valley physics, opening a new avenue for investigating geometric transport phenomena in quantum materials.

\textit{Phonon-induced geometry for Dirac fermions in graphene.—}
We consider graphene close to charge neutrality as an explicit representative of the class of (2+1)D Dirac materials and present its effective field theory formulation which induces a phonon Hall viscosity arising from both emergent gauge field and vielbein couplings of acoustic phonons to (massive) Dirac fermions in the vicinity of graphene's energy band valleys within the Brillouin zone. More specifically, pristine graphene at charge neutrality is a semimetal exhibiting two inequivalent band touching points in its Brillouin zone which define two valleys. A Dirac mass for graphene may be induced by explicitly breaking the inversion symmetry of the underlying honeycomb lattice by, e. g., placing it on a substrate such as hBN \cite{jung2017moire}, whereas Dirac fermions in other Dirac materials such as group-VI TMD's are naturally massive (see table I in \cite{xiao2012coupled}). As a simplified model we consider a Semenoff semiconductor parametrized by a constant and finite mass $m$ of opposite signs in the two valleys coupled to a strong background magnetic field. A valley-odd gauge field and a dynamical vielbein field represent those acoustic phonon couplings relevant to emergent or phonon Hall viscosity production \cite{selch2026emergent}. We regard moir\'e potentials for graphene on hBN or otherwise induced potentials as well as further phonon coupling mechanisms as small negligible perturbations in going forward.\par
In the vicinity of their would be Fermi points the low energy field theory action of Semenoff semiconducting graphene at chemical potential $\mu$ with mass parameter $m$ coupled to electromagnetic and acoustic phonon gauge and vielbein fields may be written in Euclidean space as
\begin{align}
\nonumber -S_E=&\int d^3xe\Psi^{\dagger}Q_E\Psi =\int d^3xe\Psi^{\dagger}(i\hbar\omega -H_E)\Psi \\
=&\int d^3xe\bar{\Psi}(\hbar\gamma_E^ae_a^{\mu}D_{\mu}+\gamma_E^0\mu +mv_F^2\tau^3)\Psi .
\label{minkowskidiracaction}
\end{align}
$\hbar$ is the reduced Planck constant. $Q_E$ and $H_E$ are the Euclidean fermion bilinear and Hamiltonian operators. The Dirac spinors are denoted by $\Psi$, $\Psi^{\dagger}$ and $\bar{\Psi}=\Psi^{\dagger}\gamma_E^0$, respectively. The Dirac spinors span sublattice/orbital ($\sigma$), valley ($\tau$) and spin spaces each of dimension two. We may therefore introduce Pauli matrices subject to each of these spaces in general. The Dirac $\gamma$-matrices refer to sublattice space and may be represented as linear combinations of Pauli matrices. While the structure in valley and spin spaces is trivial for semimetallic graphene, a Semenoff mass may be represented as a valley-odd parameter. The emergent geometric couplings of acoustic phonons are represented in the form of the vielbein field $e_a^{\mu}$, its inverse determinant $e=\det^{-1}(e_a^{\mu})$ and a valley-odd gauge field $A^{ph}_{\mu}$ within the covariant derivative $D_{\mu}=\partial_{\mu}+iA_{\mu}+i\tau^3A^{ph}_{\mu}$. The valley-even electromagnetic gauge field $A_{\mu}$ is assumed to be induced by a constant background magnetic field.\par
The coupling of phonons to electrons is microscopically described by lattice distortions parametrized by a displacement field $(u_i(x),h(x))^T$ $(i=1,2)$ with in-plane components $(u_1(x),u_2(x))^T$ and out-of-plane component $h$. The relevant quantities are the associated (symmetric) strain fields
\begin{align}
\bar{u}_{ij}=\frac{1}{2}(\partial_iu_j+\partial_ju_i+\partial_ih\partial_jh),\,\,\,\,u_{ij}=\bar{u}_{ij}(h=0).
\label{fullstrainfield}
\end{align}
Our subsequent analysis is based on results presented in \cite{selch2026emergent} strictly valid for distortions of pristine graphene and therefore relies on the assumption that graphene as part of a composite structure is well approximated by this idealistic state. We limit ourselves to in-plane strain due to small perturbative electron-flexural phonon coupling ($|\partial_ih\ll 1|$) in pristine graphene (but possibly not necessarily for graphene as part of a heterostructure \cite{sadeghi2023tunable}). We may furthermore neglect dilation effects induced by strain which are irrelevant for Hall viscous transport at leading order. This leads to the further simplification $u_{ii}\,$=$\,0$ implying the tracelessness of the strain tensor and vielbein determinant $e\,$=$\,1$ within our approximations. In-plane homogeneous strain induces the following perturbative modifications of the vielbein and associated metric fields relative to flat Euclidean space geometry 
\begin{align}
&e_a^i=v_F\Big(\delta_a^i+(1-\beta )\delta_a^ju_{ij}\Big),\label{strainvielbein}\\
&g^{ij}=e_a^i\delta^{ab}e_b^j,\,\,\,\,g_{ij}g^{jk}=\delta_i^k,\label{straininversemetric}\\
&g_{ij}=\frac{1}{v_F^2}\Big(\delta_{ij}+2(\beta -1)u_{ij}\Big)\label{strainmetric}
\end{align}
with $i,j,a=1,2$ and 
\begin{align}
e^\mu_0 =\delta^\mu_0\label{temporalvielbein}
\end{align}
at linear order in strain fields. The Fermi velocity is denoted by $v_F$. The emergent strain induced gauge field is parametrized by 
\begin{align}
A_i^{ph}=\frac{\beta}{2a}K_{ijk}\epsilon_{kl} u_{jl}.
\label{straingaugefields}
\end{align}
The lattice symmetry enters via the coefficients $K_{ijk}$ with non-vanishing components given by $K_{111}=-K_{122}=-K_{212}=-K_{221}=1$. We employ summation over repeated coordinate and Lorentz indices.
The parameter $a$ represents the primitive lattice constant and $\beta$ the Grüneisen parameter of graphene. The latter characterizes the change in hopping strength $t$ when adjacent atoms experience a nonzero relative displacement.

\textit{(Emergent) phonon Hall viscosity.—} 
We consider field theory descriptions of fundamentally lattice regularized systems with emergent continuous symmetries which are rotational and translational invariance modulo the gauge field $A_{\mu}$ giving rise to the background magnetic field.
The strain stress tensor is defined via variation by
\begin{align}
T^{s}_{ij}(x)=-\hbar\frac{\delta\log Z}{\delta u^{ij}(x)}=-\hbar\frac{1}{Z}\frac{\delta Z}{\delta u^{ij}(x)}.
\end{align}
with effective Euclidean partition function $Z\,$=$\,Z(e_a^{\mu}(u_{ij}),A^{ph}_{\mu}(u_{ij}),A_{\mu})$. Viscosity induces stress via strain rates
\begin{align}
T^s_{ij}=-\eta_{ijkl}\partial_0u_{kl}+... 
\end{align}
in form of the viscosity tensor $\eta_{ijkl}$. Under the assumed symmetries only one time-reversal odd viscosity component remains which is the searched for Hall viscosity. A derivation of the emergent phonon Hall viscosity from the Euclidean partition function within linear response theory involves a two-time application of the variational chain rule via the implicit dependence of the emergent geometry on the strain fields. This relates it to the Hall conductivity (within the so-called electronic Hall viscosity) and the Hall viscosity (referred to as the geometric Hall viscosity) of the electron Dirac fluid. The result for the emergent Hall viscosity is \cite{selch2026emergent}
\begin{align}
\nonumber \eta^{em}_H=&\frac{1}{4}\epsilon^{ik}\delta^{jl}\frac{\partial \bar{T}_{ij}^s}{\partial (\partial_0u_{kl})}
=\frac{\hbar}{2\pi}\Big[\Big(\frac{\beta}{2a}\Big)^2\mathcal{N}_{\sigma}+\\
&\frac{\beta (1-\beta )}{2al_B}(\mathcal{M}_1+\mathcal{M}_2)+\frac{(1-\beta )^2}{l_B^2}\mathcal{N}_{\eta}\Big]
\label{topologicalemergenthallviscosity}
\end{align}
with magnetic length $l_B=\sqrt{\frac{\hbar}{|eB|}}$ and electron charge quantum $e$ for an out-of-plane background magnetic field $B$ and a bar $\bar{(\cdot )}$ indicating a configuration space average. The calligraphic coefficient $\mathcal{N}_{\sigma}$ represents the Hall conductivity Chern number related to the electron Hall conductivity by $\sigma_H=\frac{1}{2\pi}\frac{e^2}{\hbar}\mathcal{N}_{\sigma}$ (AA-response). $\mathcal{N}_{\eta}=\frac{1}{4}\mathcal{N}_{\sigma}\mathcal{S}$ represents the Hall viscosity per density ratio with Wen-Zee shift $\mathcal{S}$ which implies an electron Hall viscosity of $\eta_H=\frac{\hbar}{2\pi l_B^2}\mathcal{N}_{\eta}$ (ee-response). The Wen-Zee shift is defined by
\begin{align}
N=\nu N_{\Phi}+\mathcal{S}\frac{\chi}{2}
\end{align}
with Euler characteristic $\chi$, the electron-like charge carrier number $N$ and the number of flux quanta $N_{\Phi}=\frac{BA}{\Phi_0}$ with flux quantum $\Phi_0=2\pi \hbar /e$. In \cite{selch2026emergent} it has been shown that $\mathcal{M}_1=\mathcal{M}_2=0$ (Ae- and eA-responses).
Note that our conventions are such that $\sigma_H=-\frac{1}{2}\epsilon^{ij}\sigma_{ij}$ as well as $\eta_H=-\frac{1}{4}\epsilon^{ik}\delta^{jl}\eta_{ijkl}$ with conductivity and viscosity tensors $\sigma_{ij}$ and $\eta_{ijkl}$ and $\epsilon^{xy}=-\epsilon^{yx}=1$, respectively.
If the first $p-1$ Landau levels above the unpaired Dirac Landau level are occupied then $\mathcal{N}_{\sigma}=p+(p-1)$ and $\mathcal{N}_{\eta}=\frac{1}{4}(p^2+(p-1)^2)$ per spin degree of freedom.\par
The final result for the emergent Hall viscosity including degeneracies and doping sign $s_p=sgn\Big(p-\frac{1}{2}\Big)$ is
\begin{align}
\nonumber\eta_H^{em}=&\eta_H^{el}+\eta_H^{geo}=\Big(\frac{\beta}{2a}\Big)^2\frac{\hbar^2}{e^2}\sigma_H+(1-\beta )^2\eta_H\\
\nonumber =&\frac{\hbar}{2\pi}\Big(\frac{\beta}{2a}\Big)^2\mathcal{N}_{\sigma}+\frac{\hbar}{2\pi l_B^2}(1-\beta )^2\mathcal{N}_{\eta}\\
=&\frac{\hbar\beta^2}{2\pi a^2}\Big(p-\frac{1}{2}\Big)+\frac{\hbar (1-\beta )^2}{4\pi l_B^2}s_p\Big(p^2+(p-1)^2\Big).
\label{calculatedemhallvis}
\end{align}
For low filling factors of order one, $a=0.246$nm for the primitive lattice constant and $\beta\approx 2$ for the Grüneisen parameter of graphene as well as magnetic fields of several Tesla (T) we find $\eta_H^{el}\gg \eta_H^{geo}$ with
\begin{align}
\frac{\eta_H^{el}}{\eta_H^{geo}}\sim \frac{l_B^2}{a^2}=\frac{\hbar}{eBa^2}\approx\frac{1.1\cdot 10^4}{B[T]}.
\label{relativesizes}
\end{align}
The large value of the electronic Hall viscosity relative to the geometric Hall viscosity supports its experimental relevance. The formal derivation of the Hall viscosity of electrons via the stress tensor uses abstract coordinate displacements which may be implemented by metric derivatives in the presence of rotational symmetry. We rely on this formalism \cite{haldane2015geometry,selch2026emergent,selch2026nonrenormalization}. The use of lattice displacements in the derivation and the absence of the electronic Hall viscosity in hydrodynamic electron flow as argued in \cite{selch2026emergent} indicates that $\eta_H^{em}$ is in fact a phonon Hall viscosity. From the derivation of the emergent Hall viscosity it can be seen that we may interpret it exactly as the coefficient of an effective acoustic phonon action obtained by integrating out (massive) Dirac electrons in the presence of a quantizing background magnetic field.\par
Apart from an emergent Hall viscosity of acoustic phonons the effective geometric couplings of phonons to electrons in Dirac materials have been shown to induce a magnetic moment for optical (degenerate) phonons with contributions being as well proportional to electron Hall conductivity and viscosity, respectively \cite{chen2025gauge,chen2025geometric}. The resulting optical phonon mass splitting into circularly polarized eigenstates may then be parametrized as a Zeeman shift at low magnetic fields. The magnetic moment of optical phonons in Dirac materials has indeed been measured via optical spectroscopy with typical sizes on the order of the electron Bohr magneton and the presented geometric formulation seems to account for it quantitatively. Gauge (vielbein) fields contribute to the induced splitting for inversion even (even and odd) optical phonon modes implying indirect sensitivity to electron Hall conductivity (viscosity). These recent findings directly motivate our studies of thermal Hall transport induced by the very same geometric formalism.

\textit{Phonon intrinsic thermal Hall transport from electron-induced phonon Berry curvature.—}
For an effective phonon action derived from integrating out electrons to be a local functional of the displacement field $u_i$ we need the electron energy eigenstates to be discrete with a chemical potential inside a gap. In graphene a considerable Landau level energy gap may be achieved for quantizing magnetic fields in the integer quantum Hall regime with small filling factor. A local acoustic phonon action is then valid if $\hbar\omega ,\hbar kv_{T,L},k_BT\ll \Delta_{LL}(p)$ with phonon frequency $\omega$, wave vector $k$, transverse (T) and longitudinal (L) velocities $v_{L,T}$, Boltzmann constant $k_B$, temperature $T$ and Landau level energy gap 
\begin{align}
\Delta_{LL}(p)=\sqrt{2\hbar eBv_F^2(1+\gamma^2)}\Big(\sqrt{p}-\sqrt{p-1}\Big)
\end{align} 
between Landau levels $p$ and $p-1\geq 0$ above the unpaired zero Landau level for n-doping with $\gamma =\frac{mv_F}{\sqrt{2\hbar eB}}$.\par
We will subsequently closely follow \cite{ye2021phonon} and represent the acoustic phonons in rotationally invariant (bulk) (2+1)D insulators by the following local effective action
\begin{align}
\nonumber S_{ph}=\frac{1}{2}\int d^2xdt(&\rho \partial_tu_i\partial_tu_i-\lambda_{ijkl}\partial_iu_j\partial_ku_l\\
&+\eta_{ijkl}\partial_iu_j\partial_k\partial_tu_l).
\end{align}
The coefficient $\rho$ represents the lattice ionic mass density of crystal atoms which derives from carbon ions for graphene. Rotational invariance implies that (i) the elasticity tensor $\lambda_{ijkl}$ as well as the viscosity tensor $\eta_{ijkl}$ are symmetric under exchange of the first and second pair of indices and (ii) the number of total independent coefficients is rather small. We may decompose the elasticity tensor with Lam\'e coefficients $\lambda$ and $\mu$ according to
\begin{align}
\lambda_{ijkl}=\lambda \delta_{ij}\delta_{kl}+\mu (\delta_{ik}\delta_{jl}+\delta_{il}\delta_{jk}).
\end{align}
These imply the transverse $v_T=\sqrt{\mu /\rho}$ and longitudinal $v_L=\sqrt{(\lambda +2\mu )/\rho }$ phonon velocities. Similarly the viscosity tensor decomposes into
\begin{align}
\nonumber \eta_{ijkl}=&\eta (\delta_{ik}\delta_{jl}+\delta_{il}\delta_{jk})+(\zeta -\frac{2}{d}\eta )\delta_{ij}\delta_{kl}\\
&-\frac{\eta_H}{2}(\epsilon_{ik}\delta_{jl}+\epsilon_{jk}\delta_{il}+\epsilon_{il}\delta_{jk}+\epsilon_{jl}\delta_{ik}).
\end{align}
We are primarily interested in and therefore only retain the non-dissipative component $\eta_H$ of the viscosity tensor which is the phonon Hall viscosity, as opposed to the dissipative bulk and shear viscosities $\zeta$ and $\eta$, respectively, and henceforth identified with $\eta_H^{em}$. Note that the condition $k_BT\ll \Delta_{LL}(p)$ mentioned above is crucial for the adopted assumption that $\eta_H$ coincides with its above presented zero temperature limit.
The local effective phonon action may be written in spatial momentum space as follows
\begin{align}
\nonumber &S_{ph}=\int \frac{dtd^2k}{(2\pi )^2}\mathcal{L}_{ph}(\bold{u}(k),\partial_t\bold{u}(k)),\\
\nonumber &\mathcal{L}_{ph}(\bold{u}(k),\partial_t\bold{u}(k))=\frac{1}{2}\Big(\rho \partial_t\bold{u}^T(-k)\partial_t\bold{u}(k)\\
&\quad\quad\,\,\,\, -\bold{u}^T(-k)\bold{M}(k)\bold{u}(k)+\partial_t\bold{u}^T(-k)\bold{A}(k)\bold{u}(k)\Big)
\end{align}
with
\begin{align}
\bold{M}(k)=\begin{pmatrix}
\lambda k_x^2+\mu (2k_x^2+k_y^2) & (\lambda +\mu )k_xk_y\\
(\lambda +\mu )k_xk_y & \lambda k_y^2 +\mu (2k_y^2+k_x^2)
\end{pmatrix}
\end{align}
and ($k^2=k_x^2+k_y^2$)
\begin{align}
\bold{A}(k)=\begin{pmatrix}
0 & \frac{\eta_H}{2}k^2\\
-\frac{\eta_H}{2}k^2 & 0
\end{pmatrix}
\end{align}
such that $\bold{M}^T(k)=\bold{M}(k)$ and $\bold{A}^T(k)=-\bold{A}(k)$, respectively. We derive the phonon Berry curvature induced by the phonon Hall viscosity at zero temperature. The so obtained Berry curvature is valid in the low temperature quantum Hall regime.\par
As a first step we derive the momentum space Hamiltonian in the first order formulation. We 
introduce the phase space wave function $\Phi (k)=(\bold{u}(k),\bold{p}(k))^T$ with displacement field $\bold{u}(k)$ and associated canonical phonon momentum
\begin{align}
\bold{p}(k)=\frac{\partial \mathcal{L}_{ph}}{\partial (\partial_t\bold{u}(-k))}=\rho \partial_t\bold{u}(k)+\frac{1}{2}\bold{A}(k)\bold{u}(k).
\end{align}
Denote by $\mathcal{H}_{ph}(\bold{u}(k),\bold{p}(k))=\frac{1}{2}\Phi^T(-k)H_{ph}\Phi (k)$ the momentum space Hamiltonian derived from $\mathcal{L}_{ph}$ by a Legendre transformation whose explicit form is given by
\begin{align}
\nonumber \mathcal{H}_{ph}&(\bold{u}(k),\bold{p}(k))=\frac{1}{2\rho}(\bold{p}(-k)-\frac{1}{2}\bold{A}(-k)\bold{u}(-k))^T\cdot\\
&(\bold{p}(k)-\frac{1}{2}\bold{A}(k)\bold{u}(k))+\frac{1}{2}\bold{u}(-k)^T\bold{M}(k)\bold{u}(k).
\end{align}
We obtain the phonon Schrödinger equation 
\begin{align}
i\hbar\partial_t\Phi (k)=H_1(k)\Phi (k),\,\,\,\,H_1(k)=iJH_{ph}(k)
\end{align}
with symplectic structure $J=\mathbbm{1}_{\bold{u}/\bold{p}}\otimes i\sigma^2_{\bold{u}\leftrightarrow \bold{p}}$ and first order Hamiltonian
\begin{align}
H_1(k)=\frac{i}{\rho}\begin{pmatrix}
-\frac{1}{2}\bold{A}(k) & \mathbbm{1} \\
-\rho \bold{M}(k) & -\frac{1}{2}\bold{A}(k)
\end{pmatrix}.
\end{align}
As a second step we outline the derivation of the left- and right-eigenvectors $l_{\sigma\tau}$ and $r_{\sigma\tau}$ ($\sigma ,\tau =\pm $) of the non-Hermitian first order Hamiltonian $H_1(k)$ normalized according to 
\begin{align}
l^T_{\sigma\tau}r_{\sigma^{\prime}\tau^{\prime}}=\delta_{\sigma\sigma^{\prime}}\delta_{\tau\tau^{\prime}}.
\end{align}
The phonon Berry curvature may be written in terms of them as
\begin{align}
\Omega_{\sigma\tau}=i\epsilon_{ij}(\partial_{k_i}l^T_{\sigma\tau})(\partial_{k_j}r_{\sigma\tau}).
\end{align} 
We note that we will tackle the eigenproblem perturbatively in the phonon Hall viscosity $\eta_H$ and retain only the leading terms which are linear in it. We define
\begin{align}
&tan(2\theta (k))=\frac{2M_{12}(k)}{M_{11}(k)-M_{22}(k)}\\
&\Delta (k)=\sqrt{(M_{11}(k)-M_{22}(k))^2+4M_{12}^2(k)}\\
&m_{\tau}=\frac{1}{2}Tr(\bold{M}(k))+\frac{\tau}{2}\Delta (k).
\end{align}
Then the energy eigenvalues of $H_1(k)$ are given by
\begin{align}
E_{\sigma\tau}=-\sigma \sqrt{\frac{m_{\tau}}{\rho}}
\end{align}
with $\bold{M}(k)\Psi_{\tau}=m_{\tau}\Psi_{\tau}$ and
\begin{align}
\Psi_+=(cos(\theta ),sin (\theta ))^T,\,\,\,\,\Psi_-=(-sin(\theta ),cos(\theta ))^T.
\end{align}
Finally the left- and right-eigenvectors of $H_1(k)$ are found to be
\begin{align}
\nonumber r_{\sigma\tau}=&
\begin{pmatrix}
\Psi_{\tau} \\
(-i)\rho E_{\sigma\tau}\Psi_{\tau}
\end{pmatrix}\\
&+\frac{1}{2}A_{12}
\begin{pmatrix}
2(-i)\frac{E_{\sigma\tau}}{\Delta}\Psi_{-\tau} \\
\frac{Tr(\bold{M})}{\Delta}\Psi_{-\tau}
\end{pmatrix}
\end{align}
and
\begin{align}
\nonumber l_{\sigma\tau}=&\frac{i}{2\rho E_{\sigma\tau}}\Bigg[
\begin{pmatrix}
(-i)\rho E_{\sigma\tau}\Psi_{\tau} \\
\Psi_{\tau}
\end{pmatrix}\\
&-\frac{1}{2}A_{12}
\begin{pmatrix}
\frac{Tr(\bold{M})}{\Delta}\Psi_{-\tau} \\
2(-i)\frac{E_{\sigma\tau}}{\Delta}\Psi_{-\tau}
\end{pmatrix}\Bigg].
\end{align}
Instead of calculating the Berry curvature directly it is useful to introduce the Berry gauge field
\begin{align}
A^i_{\sigma\tau}=il^T_{\sigma\tau}\partial_{k_i}r_{\sigma\tau}
\end{align}
in terms of which the Berry curvature reads
\begin{align}
\Omega_{\sigma\tau}=\epsilon_{ij}\partial_{k_i}A^j_{\sigma\tau}.
\end{align}
We find for the Berry gauge potential
\begin{align}
A^i_{\sigma\tau}=\frac{i}{2}\partial_{k_i}ln(-iE_{\sigma\tau})-\frac{1}{2}A_{12}\frac{3m_{\tau}+m_{-\tau}}{\Delta\rho E_{\sigma\tau}}\tau \partial_{k_i}\theta .
\label{berrygaugefield}
\end{align}
The first term in Eq. (\ref{berrygaugefield}) is a pure gauge term and does not contribute to the Berry curvature. The second term gives rise to a Berry curvature contribution linear in the phonon Hall viscosity. Notice that the first order formalism implies a doubling of degrees of freedom which is trivial in the sense that $E_{\tau}\equiv E_{-\tau}=-E_{+\tau}$ as well as $\Omega_{\tau}\equiv \Omega_{-\tau}=-\Omega_{+\tau}$. We focus only on the physical positive energy $E_{\tau}$ and corresponding Berry curvature $\Omega_{\tau}$ branches and note that $\tau =+$ corresponds to transverse (T) and $\tau =-$ to longitudinal (L) acoustic modes, respectively, which we will also employ as subscripts subsequently. With
\begin{align}
\partial_{k_i}\theta (k)=-\epsilon_{ij}\frac{k_j}{k^2}
\end{align}
and
\begin{align}
E_{\tau}=\sqrt{\frac{\lambda +3\mu -\tau (\lambda +\mu)}{2\rho}}k
\end{align}
implying $E_T=v_Tk$ as well as $E_L=v_Lk$ we obtain
\begin{align}
\Omega_{\tau}=-\tau \frac{\eta_H}{2\rho}\frac{2(v_T^2+v_L^2)-\tau (v_L^2-v_T^2)}{(v_L^2-v_T^2)\sqrt{2}\sqrt{(v_L^2+v_T^2)-\tau (v_L^2-v_T^2)}}k^{-1}.
\end{align}
This form leads to the transverse and longitudinal contributions
\begin{align}
&\Omega_T=-\frac{\eta_H}{4\rho}\frac{v_L^2+3v_T^2}{(v_L^2-v_T^2)v_T}k^{-1},\\
&\Omega_L=\frac{\eta_H}{4\rho}\frac{3v_L^2+v_T^2}{(v_L^2-v_T^2)v_L}k^{-1}.
\end{align}
For subsequent convenience we introduce the linear combination $C(k)=\frac{1}{v_T}\Omega_T(k)+\frac{1}{v_L}\Omega_L(k)\equiv C_0k^{-1}$ with
\begin{align}
C_0=\frac{\eta_H}{2\rho}\Big(\frac{1}{v_L^2}+\frac{1}{v_T^2}\Big).
\end{align}
In a third and final step we calculate the two dimensional phonon thermal Hall conductivity which we denote by $\kappa^{ph}_H=-\frac{1}{2}\epsilon_{ij}\kappa^{ph}_{ij}$. From the calculated phonon Berry curvature components we may express it according to \cite{dhakal2025theory}
\begin{align}
\frac{\kappa^{ph}_H}{T}=\frac{3}{2\pi^3}\kappa_Q\sum_{L,T}\int_{BZ}d^2kc_2(n_B(E_{L,T}(k)))\Omega_{L,T}(k)
\end{align}
where $\kappa_Q=\frac{\pi k_B^2}{6\hbar}$ is the thermal conductance quantum, $n_B$ is the Bose-Einstein distribution function,
\begin{align}
c_2(x)=\int_0^xln^2\Big(\frac{1+t}{t}\Big)dt
\end{align}
and integration is performed over the Brillouin zone (BZ). For temperatures much lower than the Debye temperature, which implies roughly $T\ll \frac{\hbar}{k_B} \frac{v_{L,T}}{a}$ with the right hand side exceeding room temperature, the acoustic phonons towards the edge of the Brillouin zone do not contribute significantly such that the Brillouin zone may be assumed to be of infinite size to a good approximation. Note that the assumption of a temperature much lower than the Debye temperature is furthermore necessary to justify linear phonon dispersion. In polar coordinates with the integral substitution $x_{L,T}=\frac{\hbar v_{L,T}k}{k_BT}$ and trivial angular integration we may express the thermal Hall conductivity of phonons compactly as
\begin{align}
\nonumber \frac{\kappa^{ph}_H}{T}=&\frac{3}{\pi^2}\kappa_Q\frac{k_B}{\hbar}TC_0\int_0^{\infty}c_2\Big(\frac{1}{e^x-1}\Big)dx\\
\nonumber =&\frac{3}{\pi^2}\zeta (3)\Gamma (4)\kappa_Q\frac{k_B}{\hbar}TC_0\\
=&\frac{3}{2\pi}\zeta (3)\frac{k_B^3}{\hbar^2}\frac{T}{\rho}\Big(\frac{1}{v_L^2}+\frac{1}{v_T^2}\Big)\eta_H.
\end{align}
The ionic mass density of carbon is given by
\begin{align}
\rho =\frac{m_{cell}}{A_{cell}}=\frac{2m_C}{\frac{\sqrt{3}}{2}a^2}
\end{align}
with unit cell mass $m_{cell}$ and area $A_{cell}$ as well as carbon atomic mass $m_C$.

\textit{Quantum Hall vs semiclassical regime.—}
The expected thermal Hall conductivity of phonons in graphene is rather small, as chirality is imprinted from electrons onto phonons by weak electron-phonon coupling $u_{ij},u_i/a\ll 1$. Therefore its detection is favorable in a phase where is does not compete with electrons meaning suppression of bulk electron transport. This happens precisely in the quantum Hall regime at strong magnetic fields and rather low temperatures. Electric and thermal transport of electrons are confined to the sample boundaries in the form of chiral edge modes. In the integer quantum Hall phases all edge modes contribute equally to electrical and thermal transport. We compare the respective sizes of thermal Hall transport of electrons and phonons. For a quantum Hall plateau with filling factor $\mathcal{N}_{\sigma}$, the electron thermal Hall conductivity is given by 
\begin{align}
\frac{\kappa_H^e}{T}=\mathcal{N}_{\sigma} \kappa_Q.
\end{align}
in the low temperature limit. This implies for the relative size of phonon and chiral edge thermal Hall conductivities
\begin{align}
\nonumber \frac{\kappa^{ph}_H}{\kappa^e_H}&=\frac{18\zeta (3)}{4\pi^3}k_B\frac{T}{\rho}\Big(\frac{1}{v_L^2}+\frac{1}{v_T^2}\Big)\Big(\frac{\beta^2}{4a^2}+\frac{(1-\beta )^2}{l_B^2}\frac{\mathcal{S}}{4}\Big)\\
&\approx \frac{9\sqrt{3}\zeta (3)}{8\pi^3}\frac{k_BT}{m_C}\Big(\frac{1}{v_L^2}+\frac{1}{v_T^2}\Big)
\label{ratioofthermaltransport}
\end{align}
where to reach the second equality we assumed
\begin{align}
\eta_H^{el}\gg \eta_H^{geo}\,\,\Leftrightarrow\,\,l_B^2\gg a^2\,\,\Leftrightarrow\,\, \frac{1.1\cdot 10^4}{B[T]}\gg 1
\end{align}
following Eq. (\ref{relativesizes}). Inserting the physical values $v_L=19.9$km/s and $v_T=12.9$km/s for the phonon velocities \cite{cong2019probing}, as well as the carbon atomic mass $m_{C}=2.00\cdot 10^{-26}$kg and the Boltzmann constant $k_B=1.38\cdot 10^{-23}$J/K $=0.086$meV/K yields $\kappa^{ph}_H/\kappa^e_H\approx 4.4\cdot 10^{-7}\,T[K]$. Therefore the relevance of the thermal Hall transport of phonons relative to that of electrons increases linearly in temperature. From the upper temperature limit
\begin{align}
&k_BT\lesssim \Delta_{LL}(1)=\sqrt{2\hbar eBv_F^2(1+\gamma^2)}\\
\nonumber &\Leftrightarrow\,\,2.4\cdot 10^{-3}T[K]\lesssim \sqrt{1+\gamma^2}\sqrt{B[T]}
\end{align}
we obtain the relation
\begin{align}
\kappa^{ph}_H/\kappa^e_H\lesssim 10^{-4}\sqrt{1+\gamma^2}\sqrt{B[T]}.
\end{align}
With $\gamma \approx 3\cdot 10^{-2}\,mv_F^2[meV]$/$\sqrt{B[T]}$ we further deduce
\begin{align}
\kappa^{ph}_H/\kappa^e_H\lesssim \Bigg\{
\begin{array}{cc}
10^{-4}\sqrt{B[T]}\,\,\,\,\,\,\,\,\,\,\,\gamma\lesssim 1\\
3\cdot 10^{-6}mv_F^2[meV]\,\,\,\,\gamma\gg 1.
\label{thermalhallestimate}
\end{array}
\end{align}
The coefficient $\gamma$ is of order one for inversion symmetry breaking gaps of several tens of meV of graphene on hBN \cite{jung2017moire,jung2015origin} and typical magnetic fields of several Tesla implying the upper estimate in Eq. (\ref{thermalhallestimate}). So even though gauge coupling of acoustic phonons gives rise to the dominant contribution in the phonon Hall viscosity, its effect is presumably undetectable in graphene as a tiny background contribution to the chiral edge thermal Hall transport which grows linearly in temperature. 
For group-VI TMD's (MX$_2$ with M=Mo,W and X=S,Se), whose gap $mv_F^2$ is on the order of eV (see table I in \cite{xiao2012coupled}), the lower estimate in Eq. (\ref{thermalhallestimate}) would be relevant.\par
We have seen that both $\kappa^e_H,\kappa^{ph}_H\propto \mathcal{N}_{\sigma}$ which means that their ratio is unaffected for the transition to the semiclassical regime ($p\gg 1$) attained by lowering the magnetic field strength. Since the Landau level gap decreases with magnetic field, the temperature regime where our formalism is valid is confined to lower temperatures disfavoring intrinsic phonon over electron thermal Hall transport even more. For $k_BT\gtrsim \Delta_{LL}(1)$ chiral edge modes cease to exist in the sense that they merge with bulk modes and heat is transported by electrons through the bulk. Assuming the approximate validity of the Wiedemann-Franz law for electron Hall transport in the bulk we expect
\begin{align}
\frac{\kappa^e_H}{\sigma_HT}=L=\frac{\pi^2}{3}\Big(\frac{k_B}{e}\Big)^2
\end{align}
with Lorenz ratio $L$. Compared to phonon thermal Hall transport we expect the same suppression as derived from Eq. (\ref{ratioofthermaltransport}), since
\begin{align}
L=\kappa_Q/ \sigma_Q=\Big(\frac{\pi k_B^2}{6\hbar}\Big)/\Big(\frac{e^2}{2\pi\hbar}\Big)
\end{align}
with thermal and electric conductance quanta $\kappa_Q$ and $\sigma_Q$, respectively, which quantify quantized transport in the quantum Hall regime. Lastly in the semiclassical regime in the presence of bulk electron transport the dominant mechanism of thermal Hall transport of phonons may very well be phonon drag induced by momentum conserving exchange within the electron-phonon two fluid system \cite{xiang2026thermal}.

\textit{Valley thermal Hall transport.—}
We may consider thermal transport of phonons with non-trivial valley character $\kappa_{vH}^{ph}$ induced by the electron fluid in graphene or more generally in (2+1)D Dirac materials through geometric coupling to phonons. The resulting time-reversal even valley thermal Hall transport of phonons will be induced by the phonon valley Hall viscosity
\begin{align}
\eta^{ph}_{vH}=\Big(\frac{\beta}{2a}\Big)^2\frac{\hbar^2}{e^2}\sigma_{vH}+(1-\beta )^2\eta^e_{vH}
\end{align}
which is nothing but the valley emergent Hall viscosity induced by the electron fluid via valley Hall \cite{sherafati2019hall}
\begin{align}
\sigma_{vH}=\frac{1}{2\pi}\frac{e^2}{\hbar}\mathcal{N}^v_{\sigma}=\frac{1}{2\pi}\frac{e^2}{\hbar}\frac{\gamma}{\sqrt{p+\gamma^2}}
\end{align}
and valley viscous Hall \cite{selch2026valley} 
\begin{align}
\eta_{vH}=\frac{\hbar}{2\pi l_B^2}\mathcal{N}^v_{\eta}=\frac{\hbar\gamma}{8\pi l_B^2}\Big(\frac{p-1}{\sqrt{p+\gamma^2}}+\frac{p}{\sqrt{p+1+\gamma^2}}\Big)
\end{align}
transport with $\gamma =\frac{mv_F}{\sqrt{2\hbar eB}}$ for n-doping with $p\geq 1$. These formulas apply for a filling of $p-1$ Landau levels above the unpaired zero Landau level and per spin degree of freedom. As discussed above the coefficient $\gamma$ is of order one in graphene on hBN and magnetic fields of several Tesla. Therefore transport signatures in graphene associated with the valley channel are expected to be comparable to those arising from the charge channel in the quantum Hall regime. With $p\sim O(1)$ the relative importance of the valley and charge channels remains roughly unmodified for Dirac materials with large Dirac masses and $\gamma\gg 1$ such as is naturally realized for group-VI TMD's.\par
Note that instead of applying an external background magnetic field alongside other stimula inducing valley inequality and thereby enabling the extraction of the valley response we may instead solely apply a constant strain gradient which corresponds to a constant valley-odd strain pseudomagnetic field according to Eq. (\ref{straingaugefields}). The so induced phonon valley thermal Hall transport is described identically after the replacement $l_{B}^{-2} \to \beta (2a)^{-1}|\epsilon_{im}K_{ijk}\epsilon_{kl} \partial_mu_{jl}|$ characterizing the relative scaling of electromagnetic and phonon gauge fields. 

\textit{Discussion and Conclusions.—}
We have analyzed phonon thermal Hall transport in graphene close to charge neutrality induced by time-reversal symmetry breaking effects of the electron fluid transferred to acoustic phonons via geometric electron-phonon couplings. The couplings are parametrized by emergent vielbein and gauge fields within a Dirac fermion formulation. The latter coupling is expected to be dominant by several orders of magnitude as compared to the former even in the quantum Hall regime. A gauge coupling which by far outweighs the small vielbein induced electron-phonon coupling considered more naturally in previous studies (see, e. g., \cite{barkeshli2012dissipationless}) inspired our detailed investigation from a theoretical perspective. Integrating out electrons in a quantizing magnetic field produces an effective phonon Hall viscosity which we identified with the emergent Hall viscosity of \cite{selch2026emergent}. The thermal Hall transport of phonons thereby arises from an intrinsic Berry curvature induced mechanism in contrast to extrinsic mechanisms arising from skew-scattering.\par
The observability of the thermal Hall transport of phonons is expected to be favored in the topologically insulating quantum Hall phase were bulk electron transport is suppressed and effectively replaced by that of chiral edge modes. This closely mimics the situation of phonon thermal Hall transport in magnetic insulators whose observation in $\alpha-$RuCl$_3$ has recently been explained by the very Hall viscosity-induced mechanism described in this work from an experimental point of view by measuring the phonon Hall viscosity directly \cite{shragai2026phonon}. Unfortunately, following our analysis, it seems that intrinsic phonon thermal Hall transport in graphene in the quantum Hall regime may not be observable. While our findings were restricted to graphene, we expect that our results apply more generically to other (2+1)D Dirac materials.
Moreover we predicted theoretically the existence of a valley thermal Hall transport of phonons calculable via valley-polarized electrons which is comparable in size to transport in the charge channel.\par
Notice that the extrinsic skew scattering mechanism may be dominant in ultraclean graphene samples, since the so induced thermal Hall conductivity scales inversely with the impurity density so long as impurities set the mean free path. It is expected to be maximal for an impurity scattering length on the order of the sample size before vanishing towards the ultraballistic limit. It seems to be worthwhile to investigate this case in detail for mesoscopic graphene devices following results in \cite{guo2021extrinsic}. Unfortunately this requires quantitative knowledge of the coupling strengths of the different coupling channels of impurities to strain fields. The case of thermal transport thereby follows patterns known from other transport phenomena like the anomalous Hall \cite{nagaosa2010anomalous} and spin Hall effects \cite{sinova2015spin}.\par
In close analogy to our discussion of acoustic phonons the electron-phonon coupling in Dirac materials via emergent frame and gauge fields may be applied to optical phonons \cite{chen2025gauge,chen2025geometric}. In this way chiral characterstics of electrons give rise to a mass splitting of the circular components of degenerate optical phonons. This effect may be understood as a Zeeman effect of phonons with a finite magnetic moment. It seems that this route is much more promising regarding experimental relevance, as phonon magnetic moments have already been observed in Dirac materials, though the identification of the responsible microscopic mechanisms seems to be still under debate. The observation of magnetic moments induced by the gauge (frame) field channel of coupling of inversion-symmetry even (even and odd) degenerate optical phonon modes in Dirac materials may allow to measure the Hall conductivity (viscosity) of electrons indirectly.






\begin{thebibliography}{10}

\vspace{-0.24cm}


\bibitem{guo2022thermal}
S. Guo, Y. Xu, R. Cheng, J. Zhou and X. Chen.
\newblock {\em The Innovation} {\bfseries 3}, 100290 (2022).

\bibitem{shragai2026phonon}
A. Shragai, E. Horsley, S. Kim, Y. J. Kim and B. J. Ramshaw.
\newblock {\em Nature} {\bfseries 652}, 1166 (2026).

\bibitem{chen2022large}
L. Chen, M.-E. Boulanger, Z.-C. Wang, F. Tafti and L. Taillefer.
\newblock {\em Proc. Natl. Acad. Sci. USA} {\bfseries 119}, e2208016199 (2022).

\bibitem{li2023the}
X. Li et al.
\newblock {\em Nat. Commun.} {\bfseries 14}, 1027 (2023).

\bibitem{jin2025discovery}
X. B. Jin et al.
\newblock {\em Phys. Rev. Lett.} {\bfseries 135}, 196302 (2025).

\bibitem{dhakal2025theory}
R. Dhakal et al.
\newblock {\ttfamily arXiv:2407.00660v4}.  

\bibitem{lefrancois2022evidence}
\'E Lefrancois et al.
\newblock {\em Phys. Rev. X} {\bfseries 12}, 021025 (2022).

\bibitem{qin2012berry}
T. Qin, J. Zhou and J. Shi.
\newblock {\em Phys. Rev. B} {\bfseries 86}, 104305 (2012).

\bibitem{saito2019berry}
T. Saito, K. Misaki, H. Ishizuka and N. Nagaosa.
\newblock {\em Nature} {\bfseries 652}, 1166 (2026).

\bibitem{ye2021phonon}
M. Ye, L. Savary and L. Balents.
\newblock{``Phonon Hall viscosity in magnetic insulators.''}
\newblock {\ttfamily arXiv:2103.04223}.

\bibitem{liu2017pseudospins}
Y. Liu, C.-S. Lian, Y. Li, Y. Xu and W. Duan.
\newblock {\em Phys. Rev. Lett.} {\bfseries 119}, 255901 (2017).

\bibitem{flebus2022charged}
B. Flebus and A. H. MacDonald.
\newblock {\em Phys. Rev. B} {\bfseries 105}, L220301 (2022).

\bibitem{guo2021extrinsic}
H. Guo and S. Sachdev.
\newblock {\em Phys. Rev. B} {\bfseries 103}, 205115 (2021).


\bibitem{sun2022large}
X.-Q. Sun, J.-Y. Chen and S. A. Kivelson.
\newblock {\em Phys. Rev. B} {\bfseries 106}, 144111 (2022).

\bibitem{guo2022resonant}
H. Guo, D. G. Joshi and S. Sachdev.
\newblock {\em Proc. Natl. Acad. Sci. USA} {\bfseries 119}, e2215141119 (2022).

\bibitem{mangeolle2022phonon}
L. Mangeolle, L. Balents and L. Savary.
\newblock {\em Phys. Rev. X} {\bfseries 12}, 041031 (2022).



\bibitem{avron1995viscosity}
J. E. Avron, R. Seiler, and P. G. Zograf.
\newblock {\em Phys. Rev. Lett.} {\bfseries 75}, 697 (1995).

\bibitem{avron1998odd}
J. E. Avron.
\newblock {\em J. Stat. Phys.} {\bfseries 92}, 543 (1998).

\bibitem{read2009non}
N. Read.
\newblock {\em Phys. Rev. B} {\bfseries 79}, 045308 (2009).

\bibitem{read2011hall}
N. Read and E. H. Rezayi.
\newblock {\em Phys. Rev. B} {\bfseries 84}, 085316 (2011).

\bibitem{hoyos2012hall}
C. Hoyos and D. T. Son.
\newblock {\em Phys. Rev. Lett.} {\bfseries 108}, 066805 (2012).

\bibitem{bradlyn2012kubo}
B. Bradlyn, M. Goldstein and N. Read.
\newblock {\em Phys. Rev. B} {\bfseries 86}, 245309 (2009).

\bibitem{hoyos2014hall}
C. Hoyos.
\newblock {\em Int. Jour. Mod. Phys. B} {\bfseries 28}, 1430007 (2014).

\bibitem{abanov2014electromagnetic}
A. G. Abanov and A. Gromov.
\newblock {\ttfamily arXiv:1401.3703}.

\bibitem{gromov2014density}
A. Gromov and A. G. Abanov.
\newblock {\em Phys. Rev. Lett.} {\bfseries 113}, 266802 (2014).

\bibitem{cho2014geometry}
G. Y. Cho, Y. You and E. Fradkin.
\newblock {\em Phys. Rev. B} {\bfseries 90}, 115139 (2014).




\bibitem{berdyugin2019measuring}
A.~I. Berdyugin et al.
\newblock {\em Science} {\bfseries 364}, 162 (2019).

\bibitem{kim2025viscous}
Kim et. al.
\newblock {\em Phys. Rev. B} {\bfseries 112}, L201404 (2025).

\bibitem{zhang2010topological}
L. Zhang, J. Ren, J.-S.Wang and B. Li.
\newblock {\em Phys. Rev. Lett.} {\bfseries 105}, 225901 (2010).

\bibitem{barkeshli2012dissipationless}
M. Barkeshli, S. B. Chung and X.-L. Qi
\newblock {\em Phys. Rev. B} {\bfseries 85}, 245107 (2012).

\bibitem{flebus2023phonon}
B. Flebus and A. H. MacDonald.
\newblock {\em Phys. Rev. Lett.} {\bfseries 131}, 236301 (2023).

\bibitem{tuegel2017hall}
T. I. Tuegel and T. L. Hughes.
\newblock{``Hall viscosity and the acoustic Faraday effect.''}
\newblock {\em Phys. Rev. B} {\bfseries 96}, 174524 (2017).
\newblock {\ttfamily arXiv:1706.02708}.

\bibitem{castroneto2009the}
A. H. Castro Neto et al.
\newblock {\em Rev. Mod. Phys.} {\bfseries 81}, 109 (2009).

\bibitem{srivastav2019universal}
S. K. Srivastav et al.
\newblock {\em Sci. Adv.} {\bfseries 5}, eaaw5798 (2019).

\bibitem{srivastav2022determination}
S. K. Srivastav et al.
\newblock {\em Nat. Commun.} {\bfseries 13}, 5185 (2022).

\bibitem{kim2013measurement}
Kim et al.
\newblock {\em Phys. Rev. Lett} {\bfseries 110}, 227402 (2013).

\bibitem{selch2026emergent}
M. Selch and M. A. Zubkov.
\newblock {\em Phys. Lett. A.} {\bfseries 595}, 132096 (2026).

\bibitem{cortijo2015hall}
A. Cortijo, Y. Ferreiro\'s, K. Landsteiner and M. A. H. Vozmediano.
\newblock {``Hall viscosity from elastic gauge fields in Dirac crystals.''}
\newblock {\ttfamily arXiv:1506.05136}.

\bibitem{heidari2019hall}
S. Heidari, A. Cortijo and R. Asgari.
\newblock{``Hall Viscosity for optical phonons.''}
\newblock {\em Phys. Rev. B} {\bfseries 100}, 165427 (2019).
\newblock {\ttfamily arXiv:1908.00313}.

\bibitem{xiao2007valley}
D. Xiao, W. Yao and Q. Niu.
\newblock {\ttfamily arXiv:0709.1274}. 

\bibitem{song2014topological}
J. C. W. Song, P. Samutpraphoot and L. S. Levitov.
\newblock {\ttfamily arXiv:1404.4019}. 



\bibitem{song2016giant}
J. C. Song and M. A. Kats.
\newblock {\em Nan. Lett.} {\bfseries 16}, 7346 (2016).

\bibitem{gorbachev2014detecting}
R. V. Gorbachev et al.
\newblock {\em Science} {\bfseries 346}, 448 (2014).

\bibitem{shimazaki2015generation}
Y. Shimazaki et al.
\newblock {\em Nat. Phys.} {\bfseries 11}, 1032 (2015).

\bibitem{yin2022tunable}
J. Yin et al.
\newblock {\em Science} {\bfseries 375}, 1398 (2022).

\bibitem{sui2015gate}
M. Sui et al.
\newblock {\em Nat. Phys.} {\bfseries 11}, 1027 (2015).

\bibitem{mak2014the}
K. F. Mak, K. L. McGill, J. Park and P. L. McEuen.
\newblock {\em Science} {\bfseries 344}, 1489 (2014).

\bibitem{selch2026valley}
M. Selch.
\newblock {\ttfamily arXiv:2607.08648}. 

\bibitem{zhai2020topological}
X. Zhai and Y. M. Blanter.
\newblock {\em Phys. Rev. B} {\bfseries 102}, 075407 (2020).

\bibitem{chen2022magnon}
Q.-H. Chen, F.-J. Huang and Y.-P. Fu.
\newblock {\em Phys. Rev. B} {\bfseries 105}, 224401 (2022).

\bibitem{jung2017moire}
J. Jung et al.
\newblock {\em Phys. Rev. B} {\bfseries 96}, 085442 (2017).

\bibitem{xiao2012coupled}
D. Xiao, G.-B. Liu, W. Feng, X. Xu and W. Yao.
\newblock {\em Phys. Rev. Lett.} {\bfseries 108}, 196802 (2012).

\bibitem{sadeghi2023tunable}
M. M. Sadeghi et al.
\newblock {\em Nature} {\bfseries 617}, 282 (2023).






\bibitem{haldane2015geometry}
F. D. M. Haldane and Y. Shen.
\newblock{\ttfamily arXiv:1512.04502}.

\bibitem{selch2026nonrenormalization}
M. Selch.
\newblock {\em Ann. Phys.} {\bfseries 494}, 170650 (2026).

\bibitem{chen2025gauge}
W. Chen et al.
\newblock{``Gauge theory of giant phonon magnetic moment in doped Dirac semimetals.''}
\newblock {\em Phys. Rev. B} {\bfseries 111}, 035126 (2025).

\bibitem{chen2025geometric}
W. Chen, X.-W. Zhang, T. Cao, S.-Z. Lin and D. Xiao.
\newblock{``Geometric origin of phonon magnetic moment in Dirac materials.''}
\newblock {\ttfamily arXiv:2505.09732}

\bibitem{cong2019probing}
X. Cong et al.
\newblock{``Probing the acoustic phonon dispersion and sound velocity of graphene by Raman spectroscopy.''}
\newblock {\em Carbon} {\bfseries 149}, 19 (2019).


\bibitem{jung2015origin}
J. Jung, A. M. DaSilva, A. H. MacDonald and S. Adam.
\newblock {\em Nat. Commun.} {\bfseries 6}, 6308 (2015).

\bibitem{xiang2026thermal}
Q. Xiang, X. Li, X. Guo, Z. Zhu and K. Behnia.
\newblock {\em Phys. Rev Lett.} {\bfseries 136}, 056303 (2026).

\bibitem{sherafati2019hall}
M. Sherafati and G. Vignale.
\newblock {\em Phys. Rev. B} {\bfseries 100}, 115421 (2019).

\bibitem{nagaosa2010anomalous}
N. Nagaosa, J. Sinova, S. Onoda, A. H. MacDonald and N. P. Ong.
\newblock {\em Rev. Mod. Phys.} {\bfseries 82}, 1539 (2010).

\bibitem{sinova2015spin}
J. Sinova, S. O. Valenzuela, J. Wunderlich, C. H. Back and T. Jungwirth.
\newblock {\em Rev. Mod. Phys.} {\bfseries 87}, 1213 (2015).


\end{thebibliography}
\end{document}